\documentclass[a4paper,oneside,11pt,DIV=14]{scrartcl}
\usepackage{graphicx}
\usepackage{float}
\usepackage{subcaption}
\usepackage{multirow}
\usepackage[a4paper, total={6in, 8in}]{geometry}
\usepackage{booktabs}
\usepackage{graphicx}
\usepackage{multicol}
\usepackage{hyperref}
\usepackage{natbib}
\usepackage[headsepline]{scrlayer-scrpage}
\addtokomafont{section}{\rmfamily\scshape}

\title{The NetMob2024 Dataset: \\
Population Density and OD Matrices \\
from Four LMIC Countries}

\author{
    Wenlan Zhang$^{1,2}$, 
    Miguel Nunez del Prado$^{1}$, 
    Vincent Gauthier$^{3}$, 
    \\Sveta Milusheva$^{1}$ \\
    \\
    $^1$The World Bank, US \\
    $^2$University College London, UK \\
    $^3$SAMOVAR, Telecom SudParis, Institut Polytechnique de Paris, FR
}
\date{September 2024}

\begin{document}

\maketitle

\begin{abstract}
The NetMob24 dataset offers a unique opportunity for researchers from a range of academic fields to access comprehensive spatiotemporal data sets spanning four countries (Colombia, Indonesia, India, and Mexico) over the course of two years (2019 and 2020). This dataset, developed in collaboration with Cuebiq (Also referred to as Spectus)\footnote{https://www.cuebiq.com/}, comprises privacy-preserving aggregated data sets derived from mobile application (app) data collected from users who have voluntarily consented to anonymous data collection for research purposes. It is our hope that this reference dataset will foster the production of new research methods and the reproducibility of research outcomes.
\end{abstract}

%
\section{Introduction}
\label{sec:introduction}
In light of the widespread adoption of mobile technology, the digital data generated from mobile devices and mobile applications provides us with the ability to examine a multitude of human behaviors, including mobility, at a previously unattainable scale. These data have the potential to facilitate the analysis of the habits of large populations at large scales, such as those of cities or countries. In certain instances, they can even supplant traditional sources of data, such as surveys in countries where data are scarce. The value of data for development and enhancement of public policies is largely untapped, as they can help fill data gaps and provide real-time and finer-scale insights~\citep{worldbank21}. While there are clear benefits to be gained from leveraging these data for the public good, there are also a number of challenges to be addressed. These include the need to ensure data privacy and security, as well as the need to develop new methodologies for data analysis. 
In the context of data for public good, several topics emerge as particularly relevant, including the use of mobile data for transportation planning, disaster recovery and epidemic response, socioeconomic analysis, and tourism. In this paper, we start out by providing an overview of the current state of research in these areas, highlighting the key findings and challenges that remain to be addressed. We then provide an overview of the Data Challenge dataset that has been made available to facilitate new research on these and other topics. We present some of the limitations of this data as well as some descriptive analysis. Finally, we highlight additional data sources that could be combined with the data to help in studying some of these topics.

With population growth in many developing countries and concentration of resources and opportunities in urban areas, many cities around the world are facing challenges in terms of transportation. The use of mobile phone data can provide valuable insights into the movement of populations in urban areas, which can be used to inform the development of transportation infrastructure and services. For example, in~\cite{Gonzalez2008}, the authors use mobile phone data to study the movement of people in cities, with the aim of improving the efficiency of public transportation systems. Different mobile phone data is used to study the movement of people in urban areas with the aim of improving the design of transportation networks including GPS traces from mobile phones~\citep{Pappalardo2015}, mobile phone traffic data~\citep{furno17}, mobile network metadata~\citep{Khodabandelou19}, and mobile phone data~\citep{bachir19, vilella20}.      
Finally,~\cite{ucar21} use mobile phone metadata to study the relationship between the use of mobile services and the development of transportation infrastructure in cities. 

Disaster recovery represents a critical domain where targeted resource allocation and response during crises is of paramount importance. The provision of a dynamic view of the situation, with the objective of enhancing situational awareness, represents a key area for consideration, as it allows for a more targeted and effective response. In this context, mobile phone data of all kinds can provide valuable insights into the movement of populations during and after disasters, as well as the impact of these movements~\cite{Yabe21,Lu12,Wang14,Bagrow11}, for example, on the demand for health services or the pre-positioning of response operations~\citep{Wang21}. In addition to mobile data, other proxies for human mobility can also provide valuable information. For instance, ~\cite{Alatrista21} and \citet{Eyre20} analyze the impact of natural disasters on purchasing behavior, with a particular focus on shifts in healthcare demand to alternative facilities. For a more detailed examination of the role played by mobile phone location data in the disaster recovery process, please refer to~\cite{Yabe22}.

In low-income countries, communicable diseases continue to have a significant impact on public health, including lower respiratory infections, diarrheal diseases, HIV/AIDS, malaria, and tuberculosis. In countries where data are limited, the use of new big data sources can inform public policy interventions aimed at reducing the mortality and morbidity rates associated with infectious diseases. As early as 2008, \citet{Gonzalez2008} began exploring the potential of mobile phone data for measuring population mobility and its subsequent application to the study of epidemics, with other researchers exploring applications to different diseases including malaria, rubella, dengue and cholera~\citep{wesolowski2012quantifying, wesolowski2015impact, wesolowski2015quantifying, bengtsson2015using, milusheva2020managing}.

More recently, the utilization of data for the purpose of managing the spread of the Coronavirus Disease 2019 (Covid-19) pandemic has become a standard practice in numerous countries. This encompasses the monitoring of individuals' geographical locations with the objective of gaining insights into mobility patterns during periods of lockdown or to facilitate disease contact tracing. CDRs were not originally designed with the intention of supporting public policy-making or enabling the government to monitor the movements of individuals \citep{milusheva2021challenges}. However, they exemplify the reuse and repurposing of data for novel purposes. In this context, mobile phone data can provide valuable feedback in quantifying the effectiveness of policies, ranging from partial curfews to strict lockdowns \citep{Oliver20}. The measurement of population density, travel patterns, and population mixing can be used to estimate population movement from mobile phone data and can also be used to improve the predictions of epidemiological models for the number of cases and geographical spread. Although both private companies and government actors have produced mobile phone applications for contact tracing, their efficacy relative to more traditional forms of contact tracing has not yet been established~\citep{Servick20}.

Some economic development and well-being metrics are now derived at scale through the lens of mobility data. Various segregation related issues~\citep{Gambetta23, Gao2024inequality} are also heavily studied with the help mobility data. More specifically in the last decade, mobile phone data has opened a new perspective by measuring and mapping poverty at country levels. For instance, the work of \cite{steele2017mapping} produces accurate, high-resolution estimates of poverty distribution in Bangladesh. Another example in Guatemala, \cite{hernandez2017estimating} use CDR data to overcome the limited fiscal and budgetary resource limitations for producing poverty estimates. \cite{njuguna2017constructing} combine mobile ownership per capita and call volume per phone  with normalized satellite nightlight data and population density, to estimate the multi-dimensional poverty index (MPI) in Rwanda. In the same line, \cite{Voukelatou20} describe the advantages and limitations when calculating well-being indicators using CDR data. \cite{pokhriyal2020estimating} highlight the use of mobile phone data for cost effective recurring poverty indicators calculation in Haiti. The work of \cite{aiken2022machine} use survey data to train machine-learning algorithms to recognize patterns of poverty in mobile phone data in Togo. \cite{Gao2024Income} employ mobile phone data for income estimation via mobility indicators, activity footprints, and travel graphs with machine learning models. Additionally, \cite{emily2022program}  use mobile phone data as an input to Machine Learning models for identifying ultra-poor households in Afghanistan.



In recent studies of the tourism industry, various methodologies have been proposed to enhance the understanding of tourist behaviors and flows. \citet{kovalcsik2022capturing} introduced a methodology aimed at comprehending tourist flows by accounting for unobserved tourists. Similarly, \cite{altin2022megastar} utilized Call Detail Record (CDR) data to analyze the different types of visits to Estonia from 2006 to 2013. Expanding on the use of CDR data, the work of \cite{xu2021towards} compares tourist mobility patterns across various cities in South Korea. \cite{grassini2021mobile} focus on analyzing the volume of tourist flows to Florence, Italy. \cite{park2023analyzing} proposed a model for segmenting tourists based on their activities using mobile phone data. In Hungary, \cite{michalko2023mobility} examined tourism dynamics between large cities and their surrounding areas using mobile phone data. Lastly, \cite{sun2021identifying} developed a methodology to differentiate tourists from locals within extensive mobile phone data sets. Together, these studies illustrate a growing trend in leveraging mobile and CDR data to refine our insights into tourist behaviors and patterns.



The use of mobile phone data has opened up new avenues for research and analysis in a variety of fields, as illustrated above. By leveraging these large-scale datasets, researchers can gain unprecedented insights into human behavior and mobility patterns. Despite the many benefits of using mobile phone data for research and analysis, there are also challenges that need to be addressed, such as ensuring the development of privacy-aware algorithms for robust methodologies in a big data context. As the field evolves, researchers and policymakers must work together to address these challenges and fully realize the potential of mobile phone data for the public good. Despite the considerable progress that has been made in the field of human mobility, a number of significant challenges remain to be addressed. In their study \cite{Pappalardo23} identify several promising avenues for future research, including the development of methods for mobility data that avoid bias, a better understanding of the diversity of travel modes, as well as a better understanding of the impact of algorithms on human mobility, and interesting new developments in the development of computational models and AI for mobility modeling~\citep{Luca21}. The Netmob 2024 Data Challenge dataset can be used to address some of these and other research areas, helping to expand knowledge in this field.

This article is organized as follows. Section~\ref{sec:datasource} reviews the data sources used to generate the netmob24 dataset. Section~\ref{sec:methodology} presents the aggregation methodology and ethical considerations used for the generation of this dataset.
Section~\ref{sec:datadescription} provides a detailed description of the datasets, and the possible anomalies present within the data. Finally, we conclude with some additional resources in section \ref{resources}.

\section{Data Source}
\label{sec:datasource}
The NetMob data challenge 2024 dataset, developed in collaboration with Cuebiq, consists of aggregated datasets that have been produced using mobile application (app) data collected from users who voluntarily provided informed consent for anonymous data collection for research purposes. Through their secure Spectus Data Clean Room platform, Cuebiq makes it possible to analyze a variety of datasets, including privacy-enhanced device location data, detected stop locations, and user trajectories across multiple countries.

\begin{figure}[H]
	\centering
	\includegraphics[width=.8\textwidth]{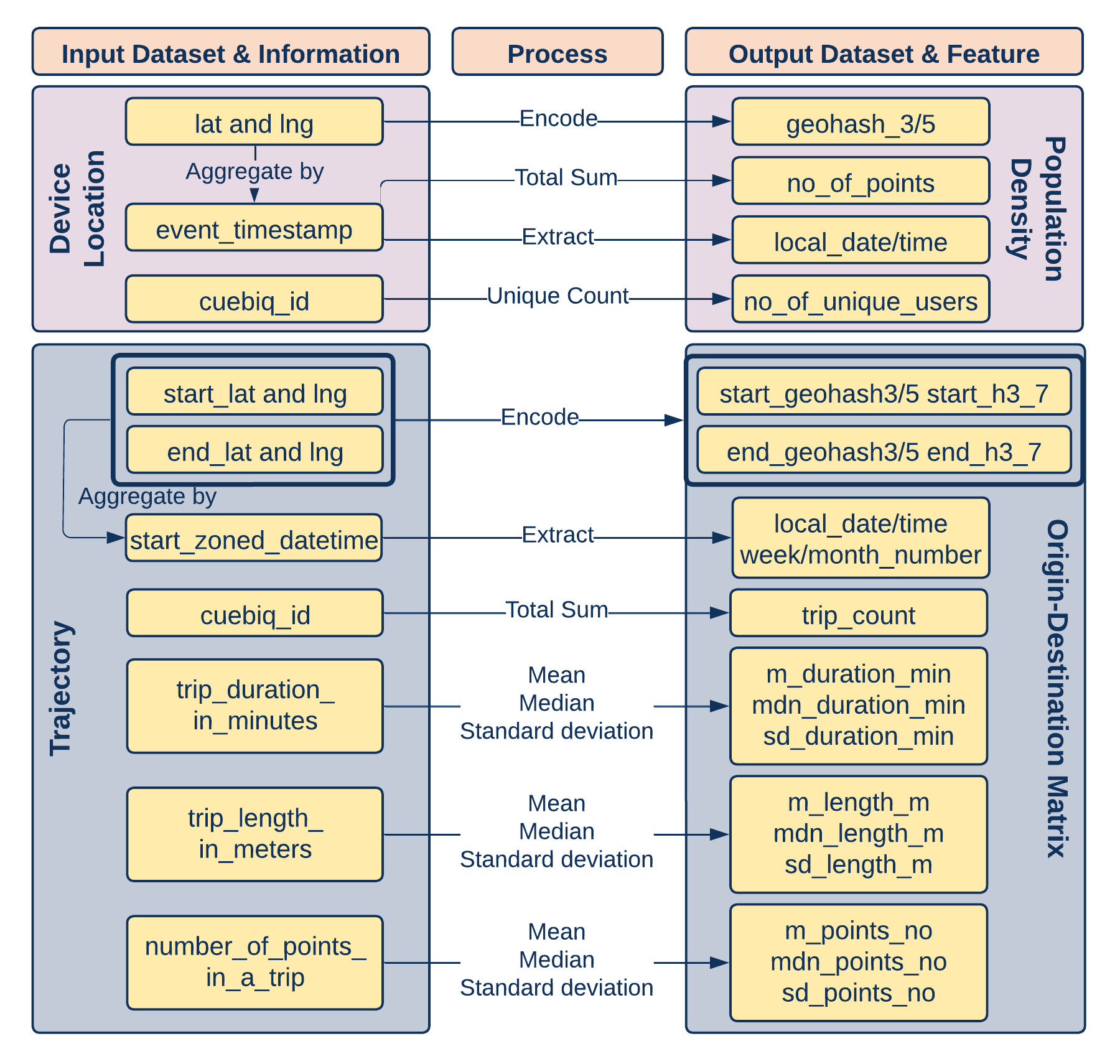}
	\caption{\small Data Processing Workflow}
	\label{fig:flowchart}
\end{figure}

Device location data capture the position of a device at a specific moment in time, recorded as individual observations. From these observations, device stop data—defined as locations where a device remained for a period—are derived using a clustering algorithm based on spatio-temporal proximity. Cuebiq also produces a trajectory dataset that includes observations on the path a device traveled between two consecutive stops within a single day. We use the  device location and trajectory datasets from four countries—Mexico, Colombia, Indonesia, and India—to prepare the data challenge datasets for Population Density (PD) and Origin-Destination (OD) Matrices, as shown in Figure \ref{fig:flowchart}. Details on the datasets and their creation follow in the subsections. These countries were chosen due to the limited existing research with mobile phone data in the context of low- and middle-income countries and data availability on the platform. The dataset covers the years 2019 and 2020, allowing for cross-year comparisons. It is important to note that data collection in Colombia began only in late October 2019, resulting in data availability for only November and December of that year.

\subsection{Device Location Data}
The device location data records the location of a given device at specific times. This dataset includes details such as the event time, anonymous ID, coordinates, accuracy in meters, operating system name, device manufacturer name, timezone offset in seconds, and speed in meters per second, among other variables. It should be noted that in some cases, the original latitude and longitude values are transformed to preserve the privacy of users. For example, home areas are re-assigned to the centroid of the corresponding Geohash 6 tile, and points falling under Sensitive Points of Interest are removed from the dataset for privacy purposes. Cuebiq first classifies the point as a recurring area, a whitelisted area\footnote{These are points of interest (POI) included in Cuebiq' POI whitelist such as commercial locations and other POIs that are permissible for use according to Cuebiq' privacy requirements.} or an "Other" area. Any point that is in a recurring area is transformed to the coordinates of the centroid of the nearest geometry with 600+ households. Importantly, the point always remains in the same geohash 6 where the original point was located.  For generating the population dataset, we use only the event time, anonymous ID, and coordinates variables. Data points with latitude or longitude values recorded as errors or zeros were removed.

\subsection{Trajectory}

The trajectory data records the path traveled between two consecutive stops by a device within a given day, capturing the movement of users throughout the day. Each trajectory provides various details, including the anonymous ID, which uniquely identifies the device without revealing personal information, and the start and end coordinates, which indicate the geographic locations where the trip began and ended. The latitude and longitude related to home locations are transformed as described above, always keeping the points within the same Geohash 6 as the original coordinates. The trajectory WKT (Well-Known Text) offers a geometric representation of the path taken.  Additional information includes the operating system name and device manufacturer, which help identify the type of device used, and the duration in minutes, which records the time spent traveling between the start and end point. The trip length in meters quantifies the distance covered, while the number of points in a trajectory indicates how many device location points were collected during the trip. It is important to note that the same user can have multiple trajectories per day, and trajectories that start on one day and end on another were filtered out in the Cuebiq trajectory data. To create the Data Challenge dataset, we focus on the anonymous ID, start and end coordinates, duration in minutes, trip length in meters, and the number of points for each trajectory.

\section{Methodology}
\label{sec:methodology}
Data has been spatially encoded and temporally aggregated to preserve personal privacy. Spatially, Geohash (GH) and H3 have been used in order to provide different levels of spatial resolution. Geohash is a widely adopted system for encoding geographic coordinates into indexes. It consists of 12 levels, with each level providing a different level of spatial precision. Each character in a geohash string corresponds to a specific level of precision, where the number of characters in the string indicates the precision level. For example, Geohash 3 (GH3), which has 3 characters, represents an area of 156 km x 156 km, while Geohash 5 (GH5), with 5 characters, represents an area of 4.9 km x 4.9 km. H3 is another global grid system for indexing geographies into a hexagonal grid, developed at Uber \citep{Uber2024}. Hexagons offer better spatial properties, such as more uniform distance between the center of the hexagon and its neighbors. This reduces edge effects and provides more accurate modeling of spatial relationships. The H3 index is represented as a 15-character (or 16-character) hexadecimal string, and the second character is a hexadecimal digit that encodes the resolution level. The selected H3 level 7 has an average edge length of 1.41km. GH3, GH5, and H37 have been selected to balance privacy concerns and maintain detailed information.

After encoding the data, all individual observations were aggregated by time interval, and the relevant features were calculated, as shown in Figure \ref{fig:flowchart}. Time aggregations include 3 hourly (3h), daily, weekly and monthly depending on the dataset. For the 3h aggregation, the time intervals are divided starting at midnight with 8 intervals during the day. For weekly data, it is important to note that the week53 from 2019 data includes only 2 days (20191230-20191231) and week 1 from 2020 includes only 5 days (20200101 - 20200105). 

\subsection{Population Density Data}

The Population Density (PD) dataset describes the presence of mobile app users, offering insights into the number of devices detected at specific locations. The dataset was generated by aggregating device location data at 3-hourly and daily intervals, using spatial units of GH3 and GH5. It includes several key attributes for analyzing spatial and temporal patterns. The \texttt{geohash\_5} or \texttt{geohash\_3} columns represent the spatial index for each observation. The \texttt{no\_of\_points} column captures the total count of observations from the device location dataset recorded within each geohash unit during a given time interval (an observation is generated every time a device has a data activity logged in one of the apps that shares data with Cuebiq), reflecting the density of data points within a specific spatial area. Note that the same device can show up multiple times in the same geohash unit and time interval, and all of these observations are summed. Additionally, the \texttt{no\_of\_unique\_users} column provides the count of distinct users, based on unique anonymous device ID in the device location dataset, associated with observations within each geohash unit for the same time interval. For this variable, a device is only counted one time for a given geohash unit and interval no matter how many device location observations it has. Finally, the \texttt{event\_time} column records the time interval of each observation, formatted as either YYYYMMDD HH:00-HH:00 for 3-hourly intervals or YYYYMMDD for daily intervals, thus providing temporal context to the data. An example of the dataset is presented in Table \ref{table:pd}

\begin{table}[ht!]
\centering
\begin{tabular}{c c c c}
\hline
\textbf{geohash\_5} & \textbf{local\_date} & \textbf{no\_of\_unique\_users} & \textbf{no\_of\_points} \\ \hline
qqg7g & 20190115 & 85 & 892 \\ 
t9rn6 & 20200102 & 10 & 86 \\ 
6rfyf & 20191103 & 88 & 940 \\ 
9u8dq & 20191230 & 95 & 1023 \\ \hline
\end{tabular}
\caption{Sample Daily Population Density Data}
\label{table:pd}
\end{table}

\subsection{Origin-Destination Matrix}

The Origin-Destination (OD) matrix dataset represents the flow of app users from a specific origin to a particular destination, providing information on the number of app user trips between different locations. The OD matrix was generated by aggregating trajectories within the same start and end spatial units of GH3, GH5 and H37 at 3-hourly, daily, weekly and monthly intervals. The dataset includes various attributes essential for analyzing travel patterns and spatial interactions. The \texttt{start\_geohash} and \texttt{end\_geohash} columns represent the spatial units for the origin and destination of trips, encoded using GH3, GH5 or H37, offering a detailed spatial reference for each trip. The \texttt{trip\_count} column aggregates the total number of trips between each start and end geohash/H3 pair for each time interval, providing insights into trip frequency. To characterize the temporal aspects of travel, the dataset includes \texttt{m\_duration\_min}, \texttt{mdn\_duration\_min}, and \texttt{sd\_duration\_min}, which respectively denote the mean, median, and standard deviation of trip durations in minutes between the start and end units. Similarly, the spatial dimensions of trips are captured through \texttt{m\_length\_m}, \texttt{mdn\_length\_m}, and \texttt{sd\_length\_m}, representing the mean, median, and standard deviation of trip lengths in meters for each day. The dataset also includes measures of observational density, with \texttt{m\_points\_no}, \texttt{mdn\_points\_no}, and \texttt{sd\_points\_no} indicating the average, median, and standard deviation of recorded device location observations per trip between geohash pairs. Same as PD, there are time columns of \texttt{local\_time} (formatted as YYYYMMDD HH:00-HH:00 for 3 hourly interval data) and 
\texttt{local\_date} (formatted as YYYYMMDD for daily data) , which records the date and datetime of each trip, providing a temporal context for the observed travel patterns. An example OD matrix is presented in Table \ref{table:od}


\begin{table}[ht!]
\hspace*{-1cm} 
\centering
\scriptsize
\begin{tabular}{c c c c c c c c c c c c c}
\hline
\multicolumn{2}{c}{\textbf{Geohash3}} & \multirow{2}{*}{\textbf{Trip No}} & \multicolumn{3}{c}{\textbf{Trip Duration (min)}} & \multicolumn{3}{c}{\textbf{Trip Length (m)}} & \multicolumn{3}{c}{\textbf{No Points per Trip}} & \multirow{2}{*}{\textbf{Date}} \\ \cline{1-2} \cline{4-12}
\textbf{Start} & \textbf{End} & & \textbf{Mean} & \textbf{Median} & \textbf{SD} & \textbf{Mean} & \textbf{Median} & \textbf{SD} & \textbf{Mean} & \textbf{Median} & \textbf{SD} & \\ \hline
6rf & 6rf & 30 & 142.48 & 48.49 & 183.83 & 17903.78 & 733.10 & 89495.21 & 4.73 & 4 & 2.79 & 20191101 \\
d0r & d0r & 141 & 78.03 & 32.15 & 129.87 & 2130.48 & 994.02 & 2837.16 & 5.34 & 4 & 4.26 & 20191101 \\
abc & def & 45 & 120.45 & 40.22 & 150.87 & 19500.56 & 750.89 & 89000.12 & 5.67 & 5 & 3.45 & 20191102 \\
ghi & jkl & 67 & 98.34 & 35.78 & 110.56 & 2500.78 & 1050.33 & 3000.45 & 6.23 & 5 & 4.12 & 20191103 \\ \hline
\end{tabular}
\caption{Sample Daily Origin-Destination Matrix Data}
\label{table:od}
\end{table}

\subsection{Ethics Considerations}

To ensure a high level of privacy for individuals, we implemented several measures while handling the mobile app data. The data is sourced only from users who have consented to share their information through various mobile applications. Before the data was shared with us, all personally identifiable information was removed, and device IDs were anonymized by Cuebiq. Additionally, Cuebiq applied a privacy enhancement method that integrates classification and transformation in a sequential process. Initially, each location point is categorized based on specific criteria. Points that fall within a "whitelisted" point of interest (POI), as defined by Cuebiq' privacy-compliant POI whitelist, or those that do not meet other classification criteria, retain their original latitude and longitude values without modification. Conversely, points identified as being near a recurring area associated with the device, likely the home location, undergo a transformation. In these cases, the original latitude and longitude values are adjusted to the centroid of the nearest geometry containing 600 or more households. The transformation ensures that the adjusted point remains within the same geohash level 6 as the original point.

\begin{figure}[H]
    \begin{subfigure}[b]{0.5\textwidth}
        \centering
        \includegraphics[width=\textwidth]{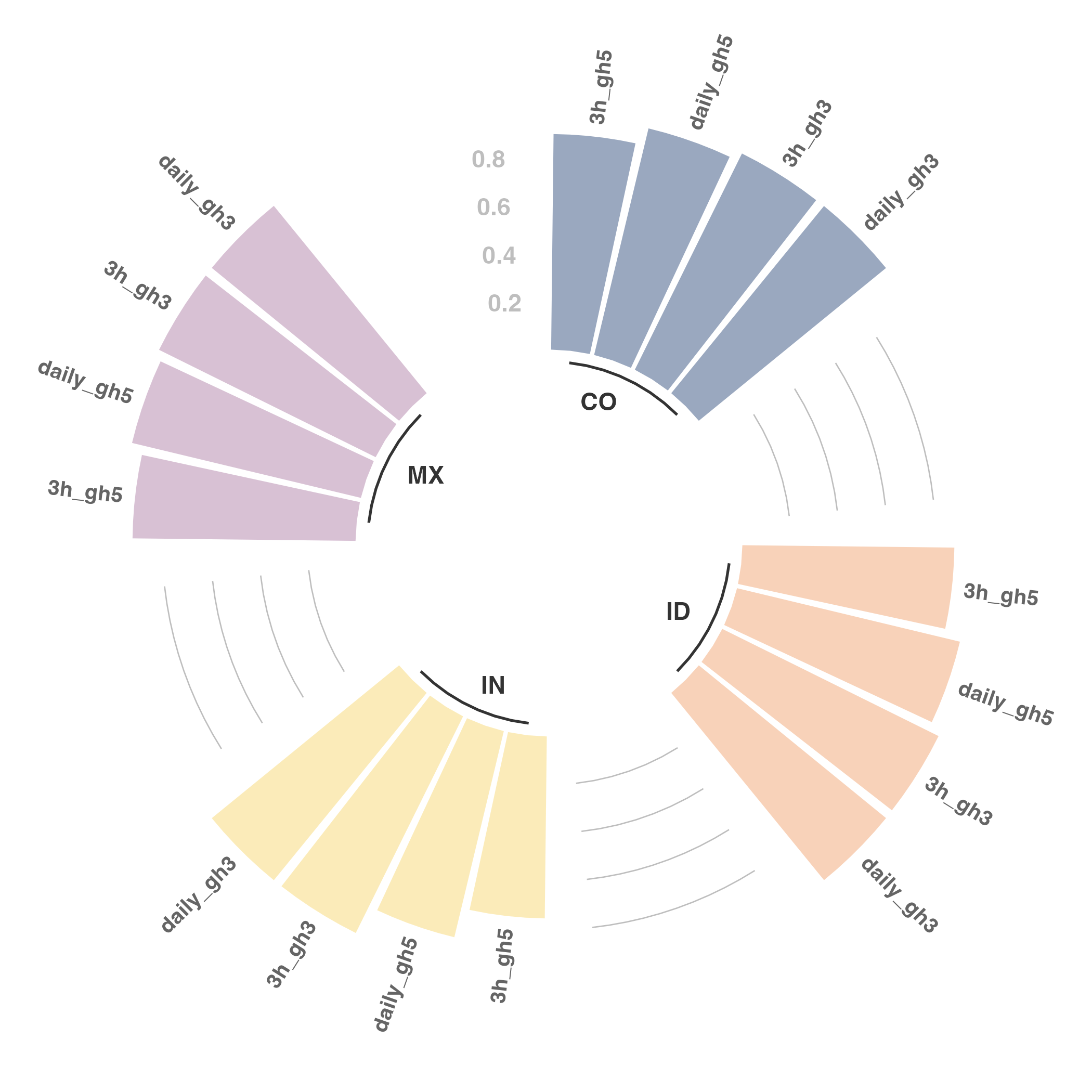} 
        \caption{\small Proportion of PD Points that are Kept compared to Total PD Points after removing cells with fewer than 10 Unique Users}
        \label{fig:kept_pd}
    \end{subfigure}
    \begin{subfigure}[b]{0.45\textwidth}
        \centering
        \includegraphics[width=\textwidth]{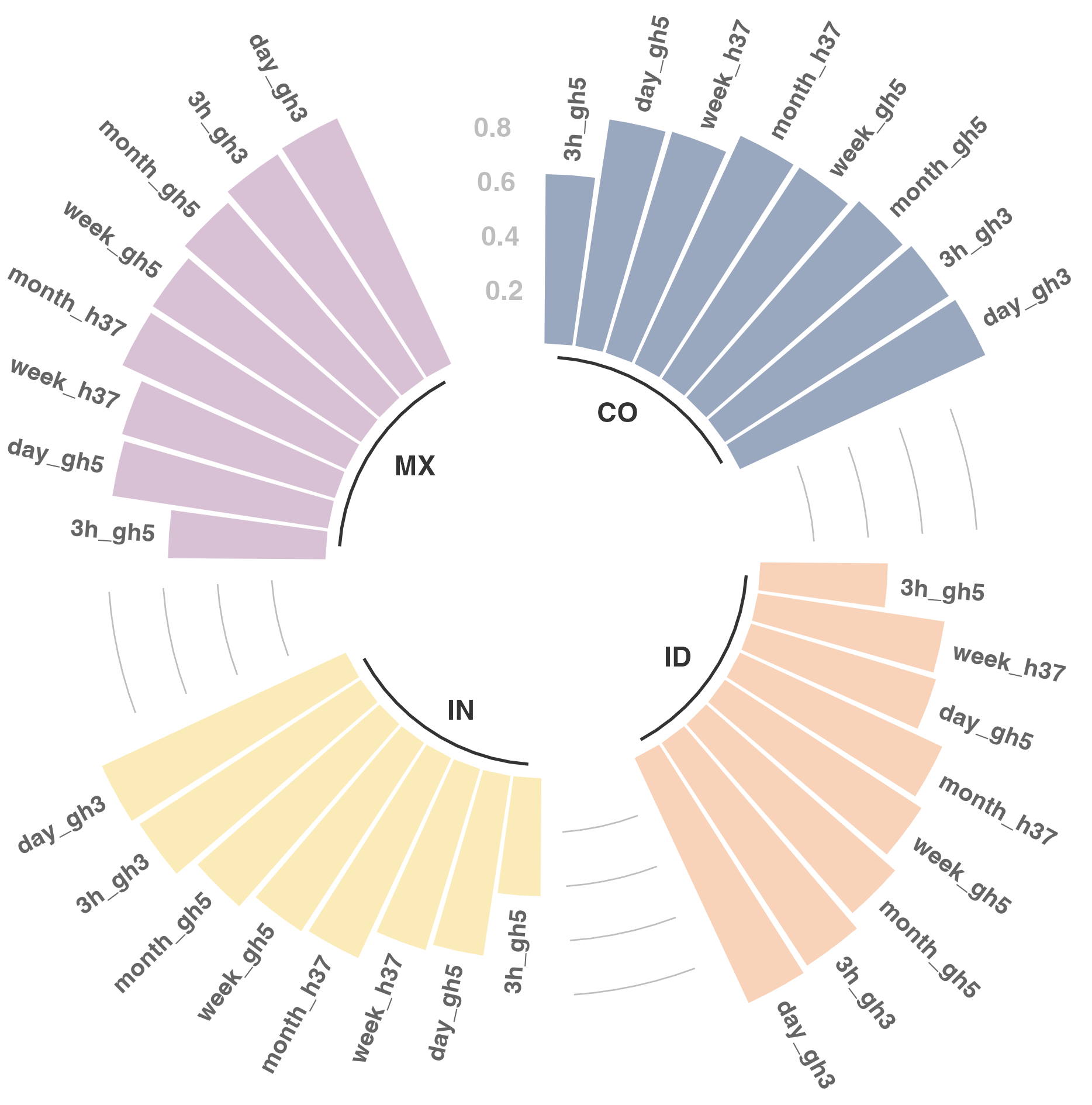} 
        \caption{\small Proportion of OD Pairs with Minimum 10 Trips that are Kept Compared to Total OD Pairs}
        \label{fig:kept_od}
    \end{subfigure}
    \caption{Preserved Data for the 4 Countries and Different Temporal and Spatial Datasets, 2019}
   \label{fig:kept_data}
\end{figure}

Upon receiving the privacy-enhanced data, the first step applied was geohash encoding using the GH3, GH5 or H37 level, followed by aggregation by 3 hourly, daily, weekly or monthly time interval, rather than individual or device-level data. Only cells with 10 or more users were included, leading to the exclusion of some cells, as illustrated in Figure \ref{fig:kept_data}. The figure shows the cell observations in the available dataset against the total cell observations from the original aggregated data, to demonstrate what proportion of data is excluded from each dataset in order to maintain the minimum of 10 users per time/location interval. All data aggregation was conducted within the secure environment of Spectus Data Clean Room. Only the aggregated and threshold-applied data were exported, with Cuebiq’s consent.

\subsection{Final dataset format}

\begin{figure}[H]
	\centering
	\includegraphics[width=1\textwidth]{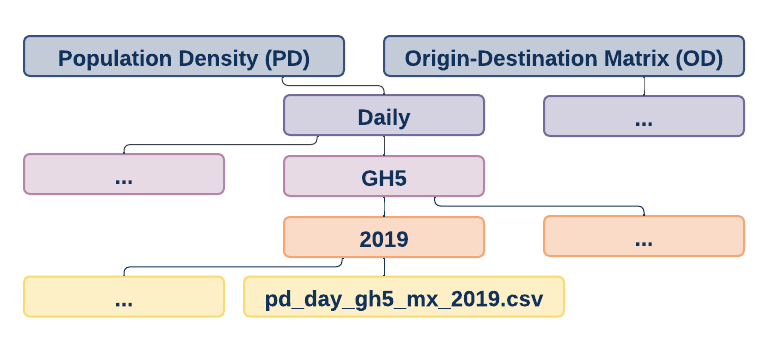}
	\caption{Hierarchical Organization of Folder}
	\label{fig:folder}
\end{figure}

The NetMob 2024 data challenge dataset includes PD and OD data from four low and middle income countries (LMICs): Colombia(CO), Indonesia(ID), India(IN), and Mexico(MX), spanning the years 2019 to 2020. Data collection for CO began in November 2019, so both PD and OD only started then. The dataset is provided based on country, spatial geoencode index and temporal interval. The hierarchical organization of files can be seen in Figure \ref{fig:folder}. The structure is based on dataset-temporal interval-encode-year-country. 




\section{Data Description}
\label{sec:datadescription}
This section presents an exploratory analysis of the provided data challenge PD and OD datasets. The analysis aims to offer an understanding of the datasets from the point of view of completeness,  spatial, and time interval perspectives, highlighting their strengths and weaknesses, to support participants with their research applications. 

\subsection{Data Completeness and Anomalies}

\begin{figure}[H]
    \begin{subfigure}[b]{0.5\textwidth}
        \centering
        \includegraphics[width=\textwidth]{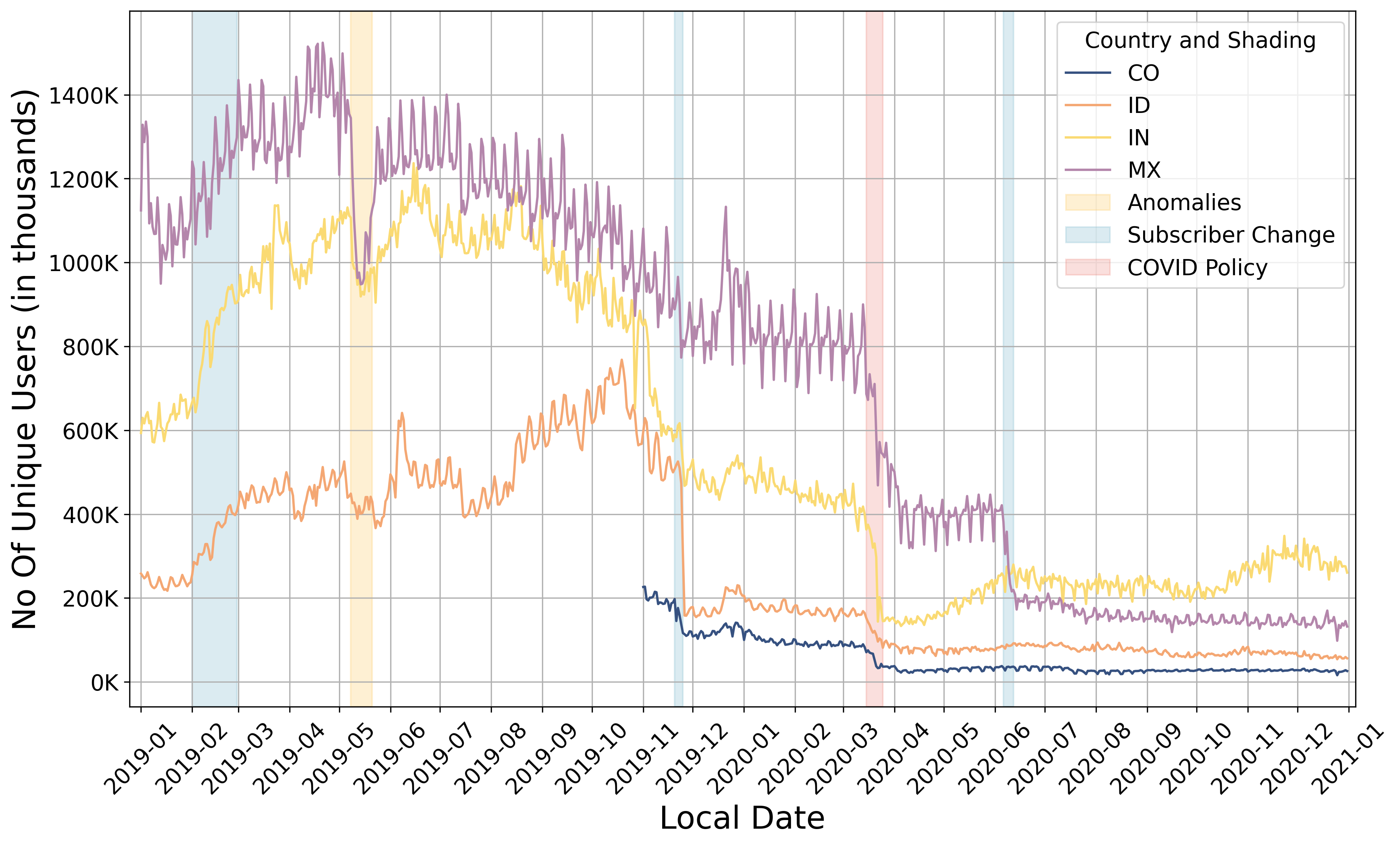}
        \caption{\small Number of Unique Users from PD}
        \label{fig:PD}
    \end{subfigure}
    \begin{subfigure}[b]{0.5\textwidth}
        \centering
        \includegraphics[width=\textwidth]{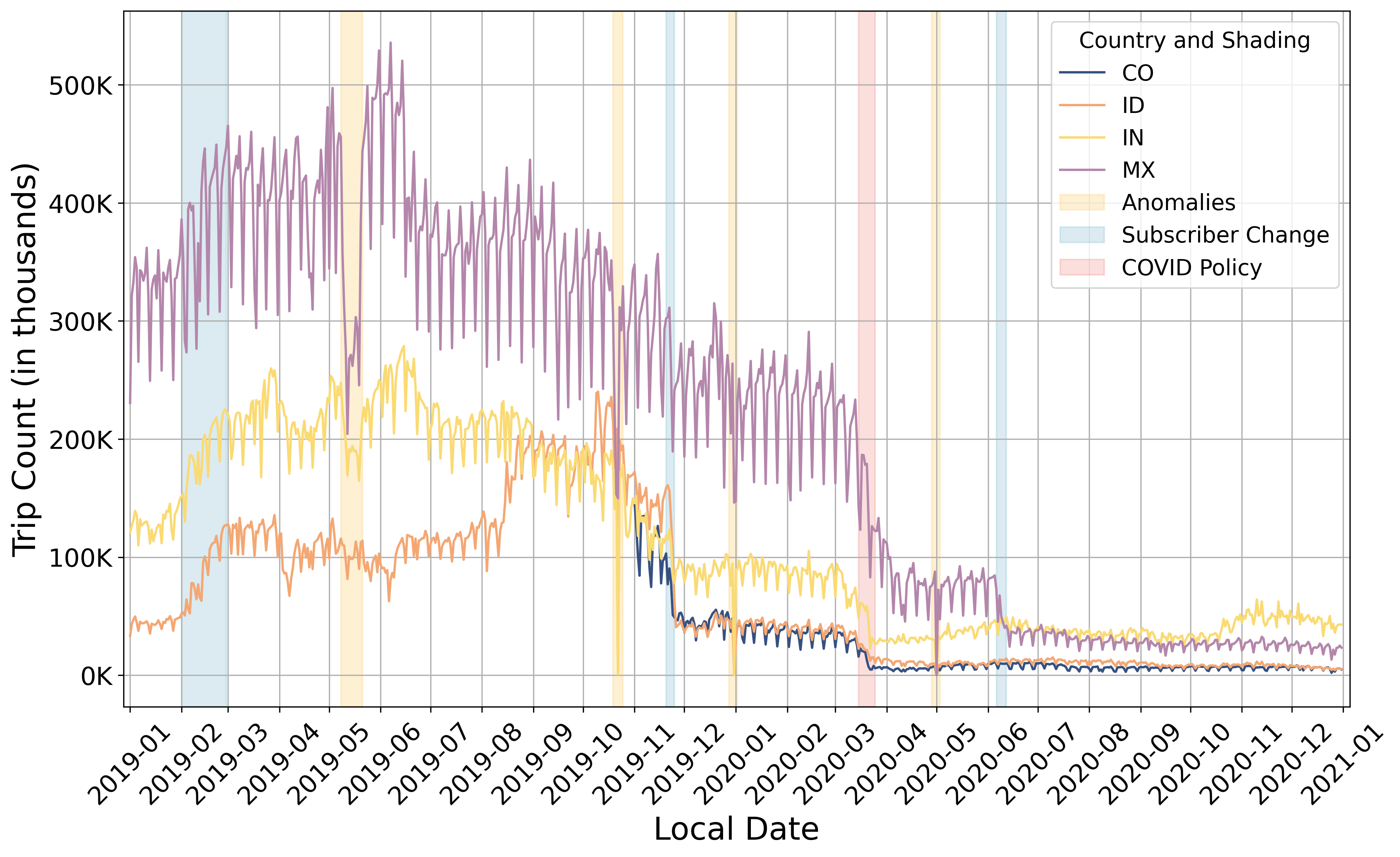} 
        \caption{\small Number of Trip Counts from OD}
        \label{fig:OD}
    \end{subfigure}
    \begin{subfigure}[b]{1.0\textwidth}
        \centering
        \includegraphics[width=0.5\textwidth]{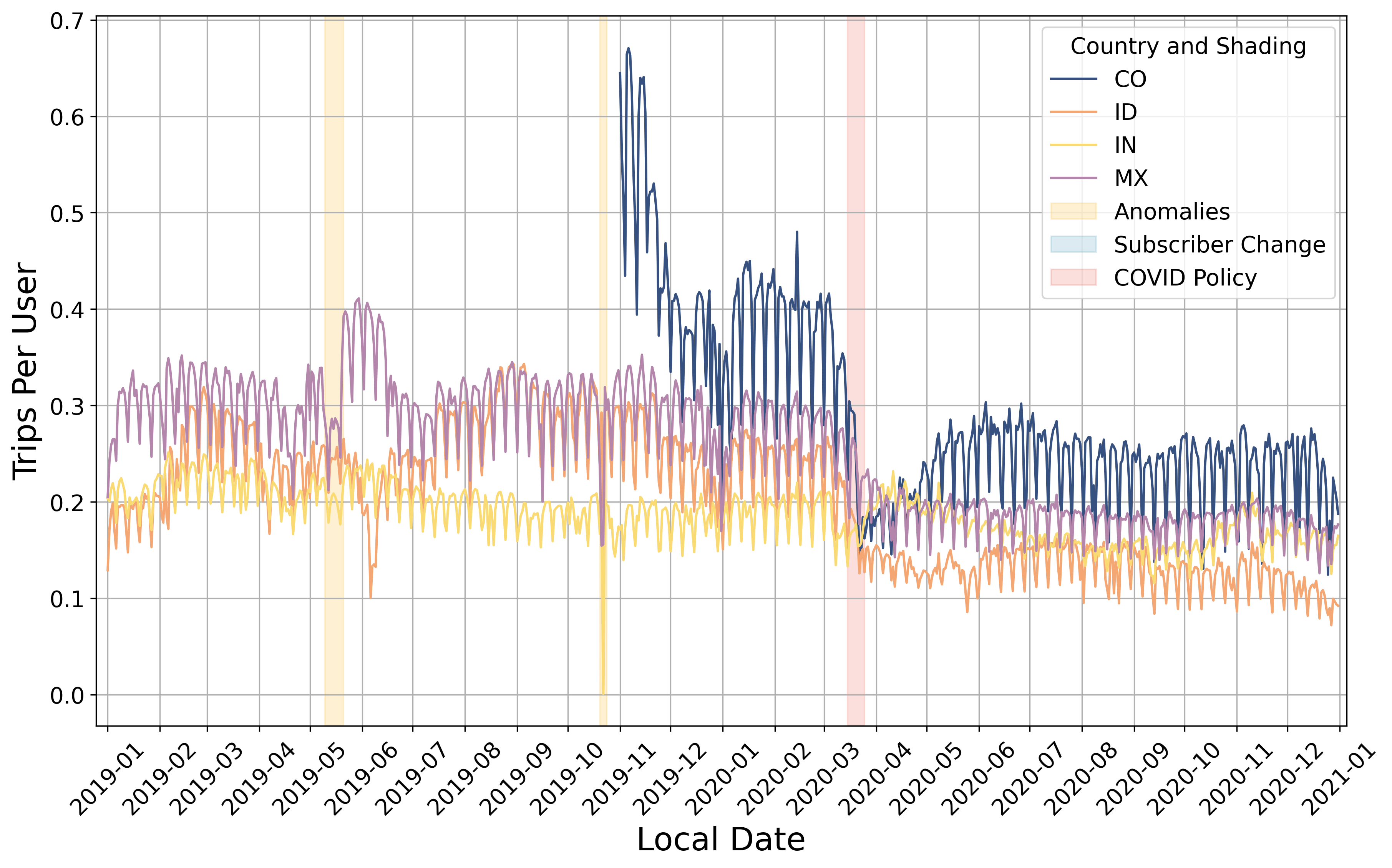} 
        \caption{\small Ratio of Trip Count to Unique Users}
        \label{fig:OD/PD_1y}
    \end{subfigure}
    \caption{\small Time Series PD, OD and OD/PD Highlighting Some Anomalies, 2019–2020}
   \label{fig:timeseries}
\end{figure}

Figure \ref{fig:timeseries} presents the time series of the number of unique users from the PD dataset, the trip count from the OD dataset, and the ratio of trip count to unique users for 2019 and 2020. This visualization aims to highlight key feature changes and identify anomalies. Visually, some of the days we can observe anomalies include May 10 to May 20, 2019, as well as on October 22, 2019. We are identifying these anomalies visually based on drops in the line charts in Figure \ref{fig:PD} and Figure \ref{fig:OD}, but when conducting analysis with the data, it may be advisable to identify all anomalies using a standard definition. An example of such a definition would be identifying observations that are more than 2 standard deviations from the average values within 30 days. It is important to note that for the OD data, certain days are missing data for certain countries. The OD has missing data for all four countries on 20200501; India and Indonesia for 20191022 and 20191231. Additionally, there are some days where certain time intervals are missing (e.g. Colombia and Mexico on 20191231 for the hourly intervals starting at 0,3,6, and 9). In the 3h datasets, these observations would be missing, but in the daily datasets, there would be an observation for the day since there are some time intervals with data, but the values would be much lower since part of the day is missing. Again, these cases could be identified through anomalie detection methods.

In addition to anomalies occurring on specific days or short periods of time, there are also important shifts in the data due to the data generating process. The data is coming from users providing their location on mobile apps. If new apps start to provide data to Cuebiq or else other apps stop providing this data, then it can lead to big shifts up or down in the number of users and the number of trips measured. We can see this happen with Indonesia in late 2019, when the number of unique users per day suddenly drops by more than half. We can also see across Mexico, India and Indonesia the ramp-up of unique users in early 2019 as more apps share their data. We do not have information on the specific apps from which the data is coming and when they start or stop sharing their data. It is important to factor in that these changes in the data generating process are occurring because otherwise they may be considered as representing real changes in the number of people in a location or the number of trips, when in fact the changes are due to a shift in the number of people providing their location.

When working with the trips dataset, one option for accounting for the change in the subscriber base contributing data or other issues in terms of anomalies in the number of observations is to also compare the proportion of trips to users. Figure \ref{fig:OD/PD_1y} demonstrates that while the number of trip counts seems to change quite significantly at different times of the year, when accounting for the number of users producing observations on a given day, the ratio stays relatively stable across the year for each country. India seems to be the most steady with a ratio around .2 trips per unique user per day. Mexico is higher, with a ratio of around .3, though there is a large increase in the ratio around June 2019 that may therefore represent a real shift in people moving more (rather than a function of more users providing data). Indonesia has more variability during the year, and Colombia starts out much higher than the other countries but then levels out close to their values. This different pattern in Colombia when the data first starts signals potentially that the first month of data may not be as reliable, therefore potentially excluding this data or conducting robustness checks with and without the data may be advisable. 

So far the focus has been on aspects of the data generating process that may impact the data at a particular time, but another important area related to completeness is the representativeness of the data. In order to be part of the sample of users in the data, it is a necessary condition to have a smartphone and to also have the resources to pay for mobile data to use the apps through which the data points are generated. When studying low- and middle-income countries, there can be large parts of the population that are not able to afford a smartphone or to pay for mobile data on a regular basis, and therefore would not be represented in the dataset \citep{milusheva2021assessing}. Additionally, the apps that different people use may differ based on demographic characteristics like gender or income, and as some apps become excluded or included in contributing to the data, the demographic make-up of those contributing their data could change, affecting also the behaviors seen in the data. This can also vary from country to country as certain apps may be more popular in some countries and not in others. These are not aspects that it is possible to correct for in the datasets provided for the Challenge, but important aspects to consider when interpreting the data and results and to be further analyzed in the future.

\subsection{Spatial and Temporal Analysis}

\subsubsection{Population Density Data}

\begin{figure}[H]
	\centering
	\includegraphics[width=1\textwidth]{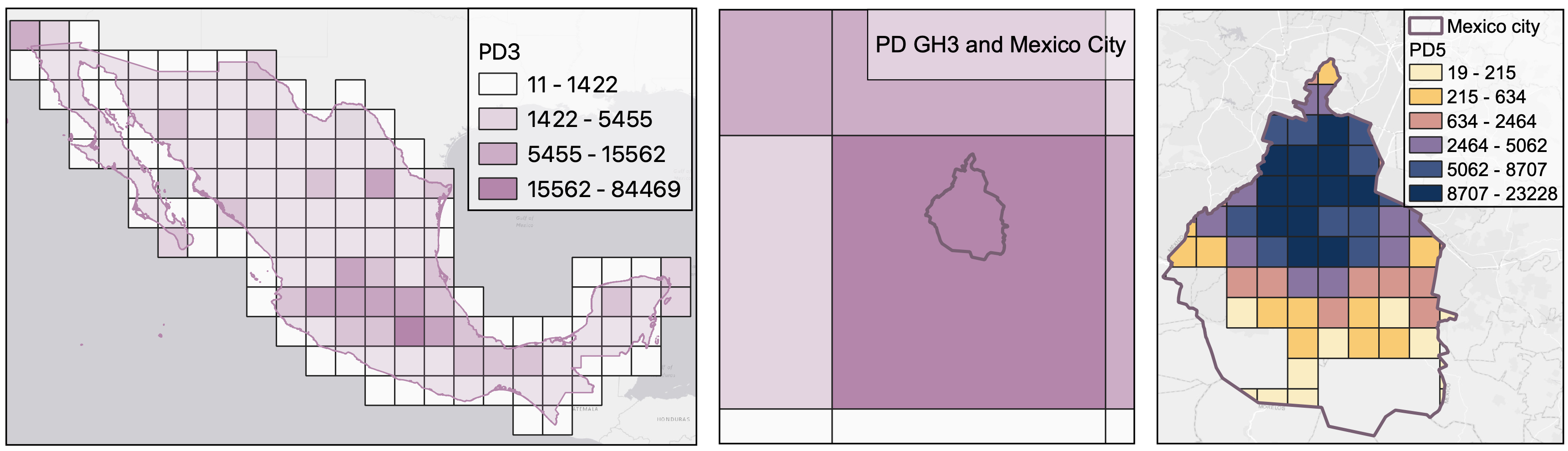}
	\caption{\small Mexico Unique Users from PD Dataset in Geohash 3 \& 5, Avg Daily Value for November 2019}
	\label{fig:PD35}
\end{figure}


Understanding population density is critical for a wide range of applications, from national-level policy planning to urban management. As shown in Figure \ref{fig:PD35}, the two levels of PD data provided can support both macro and micro-level studies.\footnote{The figure uses \cite{Jenks1967} natural breaks method for grouping cells by population density. } GH3, with its broad spatial coverage, is ideal for strategic planning and resource allocation on a macro scale, enabling policymakers to assess regional disparities or optimize large-scale infrastructure projects. In contrast, GH5 provides a finer resolution suited for city-level urban planning and management. It allows for detailed mapping of population concentrations within cities, guiding decisions on zoning, public service distribution, and disaster preparedness. Together, these geohash levels enable a comprehensive approach to population density analysis, supporting both high-level policy decisions and granular urban management. Example maps for all four countries are included in the appendix. 

\subsubsection{Origin-Destination}

It is possible to visualize some of the patterns of movement between origin and destination locations. Looking at India and Mexico, they have very different patterns of movement between origin-destination pairs. Figure \ref{fig:heatmaps} is a heatmap showing movement between the top 30 geohash 5 areas, and they have been ordered such that being closer together on the axis means they are closer together geographically. In India, focusing on the top 30 pairs in terms of total trips, they are very concentrated geographically. Mexico, on the other hand, has concentration as well (which can be seen with more blue colors along the diagonal which represents cells that are closer together geographically), but there is also movement to places further away.  

\begin{figure}[H]
    \centering
    \begin{subfigure}[b]{0.45\textwidth}
        \centering
        \includegraphics[width=\textwidth]{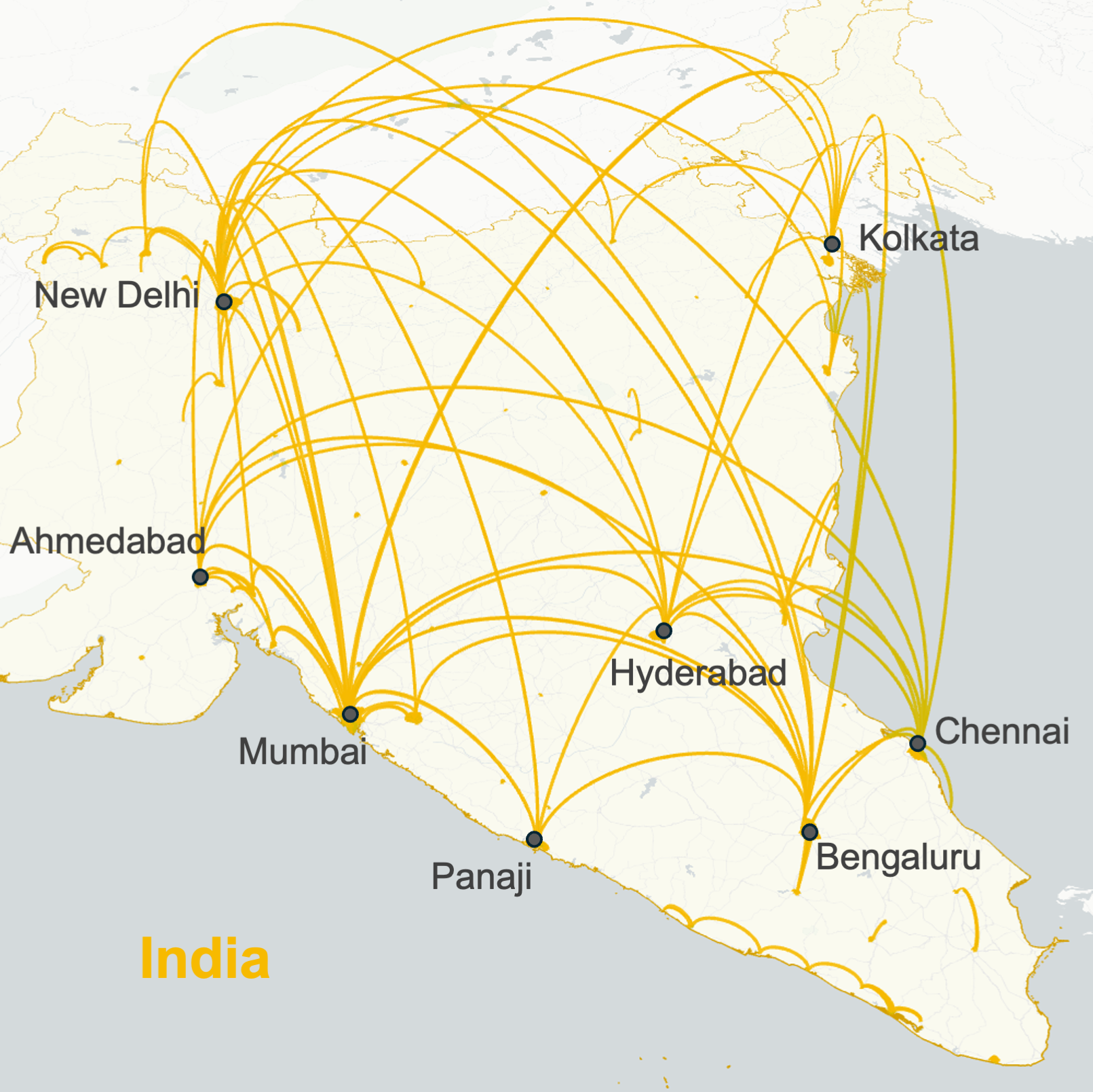} 
        \caption{\small India OD}
        \label{fig:inod}
    \end{subfigure}
    \begin{subfigure}[b]{0.45\textwidth}
        \centering
        \includegraphics[width=\textwidth]{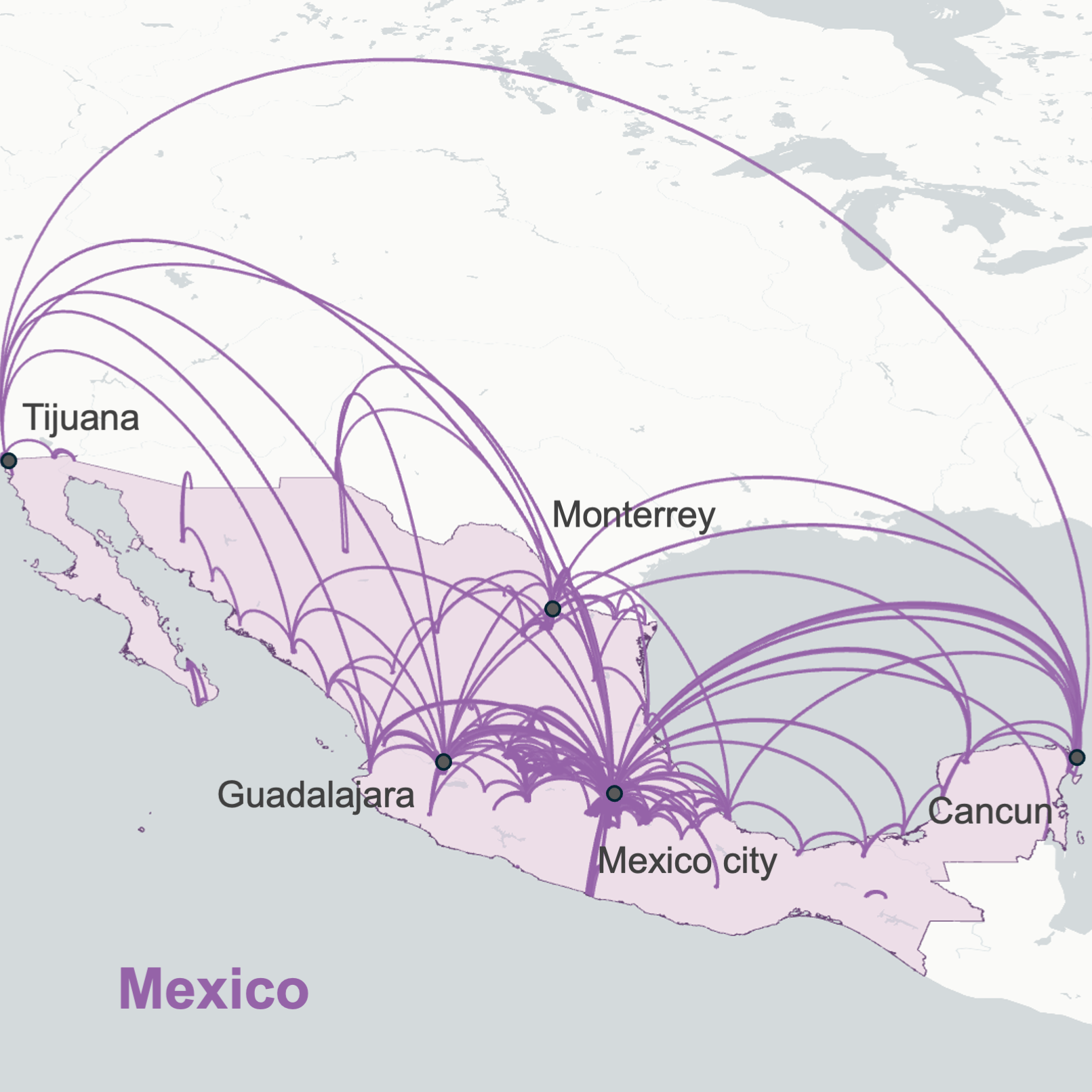} 
        \caption{\small Mexico OD}
        \label{fig:mxod}
    \end{subfigure}
    \caption{Visualization of Movement Between Different Areas of the Country, December 2019}
    \label{fig:movemaps}
\end{figure}

In Figure \ref{fig:movemaps}, we see that there is also a different pattern of movement when it comes to cross-country movement. In India, there are several major cities that act as hubs, and the concentrated movement we saw in the OD heatmap showed that high levels of movement are concentrated within those major city hubs. There is also movement that happens across the big hubs though, but it is much less than within the proximate area to the hubs. In Mexico, on the other hand, Mexico City seems to act as the main central point from which movement across the country radiates. This type of pattern was also seen in the heatmap.


\begin{figure}[H]
    \centering
    \begin{subfigure}[b]{0.45\textwidth}
        \centering
        \includegraphics[width=\textwidth]{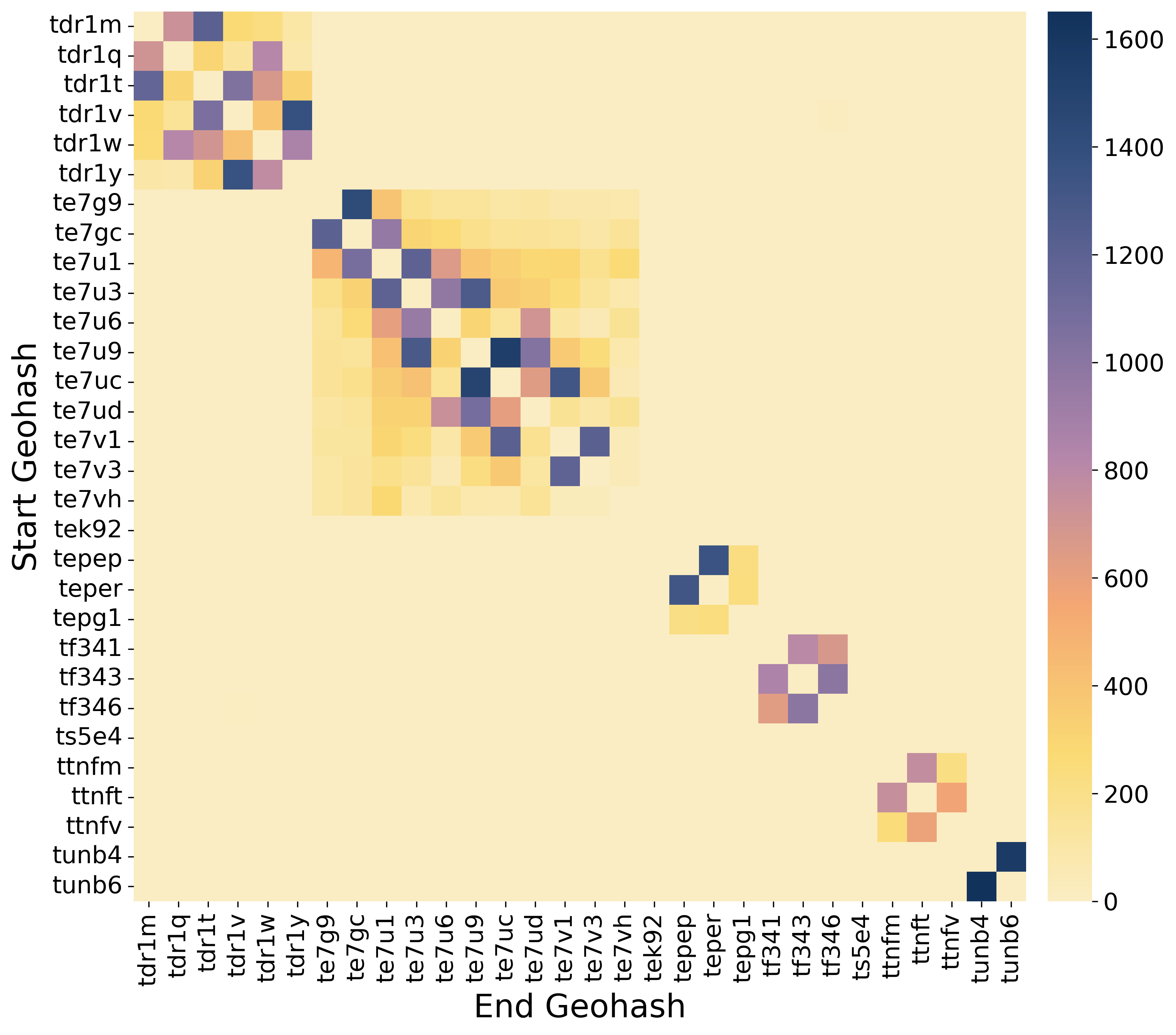}
        \caption{\small India OD}
        \label{fig:heatmap_in}
    \end{subfigure}
    \begin{subfigure}[b]{0.45\textwidth}
        \centering
        \includegraphics[width=\textwidth]{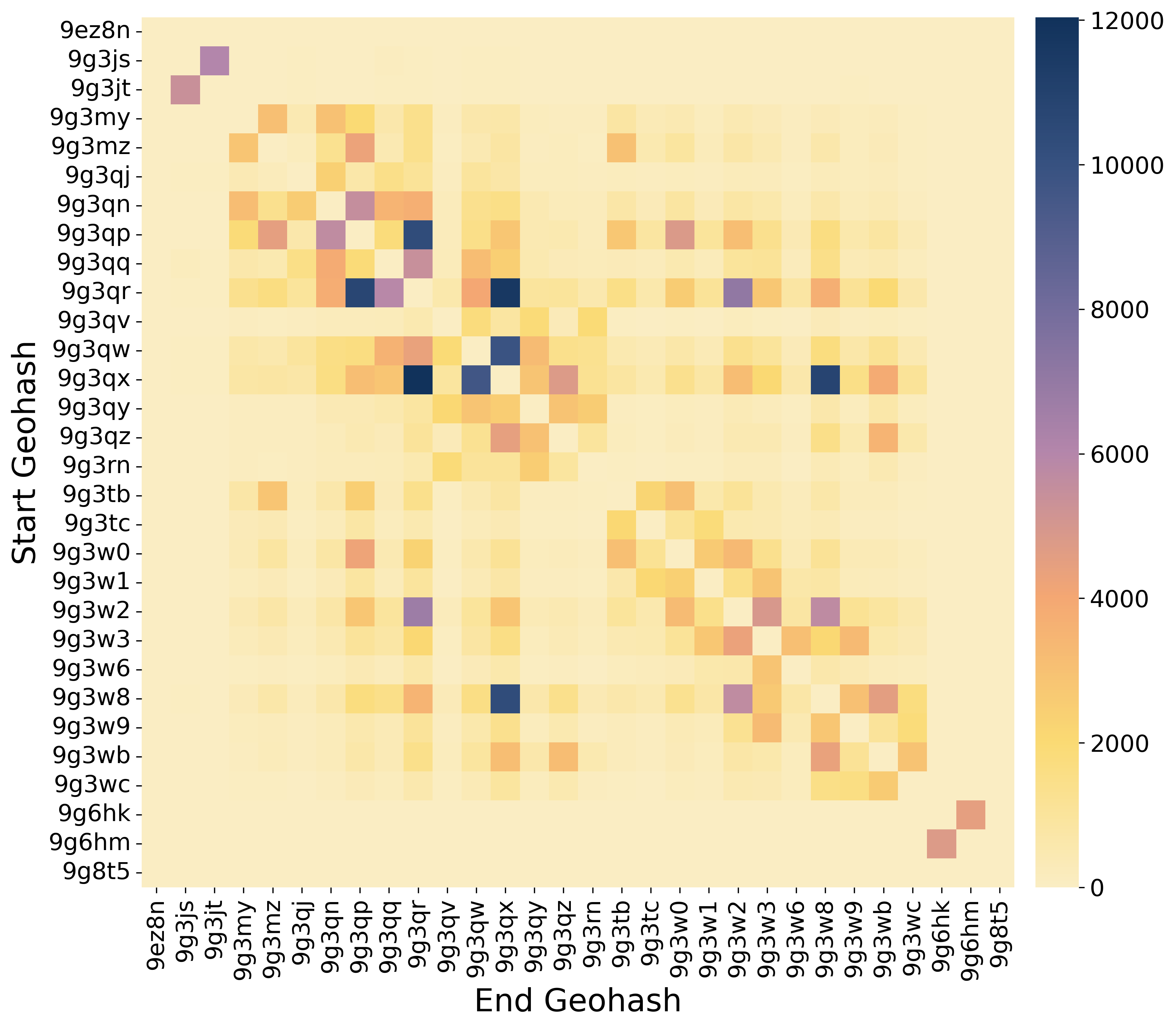} 
        \caption{\small Mexico OD}
        \label{fig:heatmap_mx}
    \end{subfigure}
    \caption{Top 30 OD Movement Geohash Pairs Heat Map, December 2019}
    \label{fig:heatmaps}
\end{figure}

\begin{figure}[H]
	\centering
	\includegraphics[width=1\textwidth]{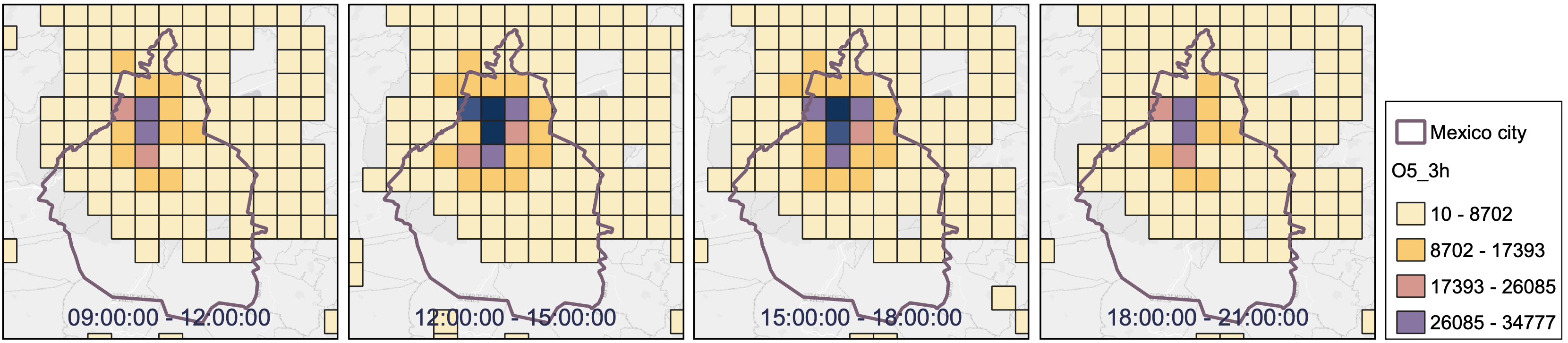}
	\caption{\small Trip Count by 3h Timely Origin in Mexico City, November 2019 at Geohash 5 Level}
	\label{fig:O5_3h}
\end{figure}

Using the 3 hour dataset, it is also possible to look at how patterns of movement change over the course of a day. Figure \ref{fig:O5_3h} shows the total trips in November across Mexico City for different time intervals. It demonstrates that most trips are detected during the middle of the day, 12:00-15:00 and 15:00-18:00 and are concentrated in the northern part of the city. Combining datasets at different time intervals and different spatial resolutions can help with learning more about the mobility patterns across the four countries. 

It is important to note that again, the data generating process may affect the final datasets. In particular, Cuebiq removes any trips that start in one day and end in a different day. Therefore, trips starting later in the day that might not end until the following day will be removed, potentially decreasing the number of trips measured in the evening. Additionally, as already discussed, the data is a function of the users providing that information. If there are parts of the city or country with a much higher proportion of low-income individuals who do not have a smartphone, are not able to afford to pay for data, or are less likely to use mobile apps that collect this data, there will be much less data collected from these areas. This is especially true for more rural areas or lower-income areas within cities.

\section{Additional Resources}
\label{resources}

Several complementary data sources have been identified that could supplement the mobile app data for the data challenge. These sources provide additional context and depth, enabling a more comprehensive analysis. They include demographic and health surveys, geospatial datasets, socio-economic indicators, and environmental data, all of which can help to validate, enrich, and cross-reference the mobile app data, offering a fuller understanding of the patterns and trends observed.

\textbf{WorldEX} \citep{Solatorio2024} is a platform that utilizes H3 indexing to facilitate the discovery of geospatial data, particularly at the sub-national level, to support socio-economic research and policy-making. The website aggregates and summarizes publicly available datasets from a variety of reputable sources, including Climate Trace, Ember Climate, the Humanitarian Data Exchange (HDX), Source Cooperative, the Uppsala Conflict Data Program (UCDP), the United Nations High Commissioner for Refugees (UNHCR), the World Bank, and WorldPop. These datasets cover a wide range of topics on a global scale, offering insights into various levels of administrative boundaries, socio-economic features such as population demographics and age distribution, natural hazards, as well as environmental factors like forest coverage and building footprints. WorldEX provides direct links to these datasets, making it a valuable resource for researchers and policymakers seeking to access and analyze detailed geospatial information for informed decision-making\footnote{\url{https://worldex.org/}}.

The \textbf{Demographic and Health Surveys (DHS) Program}, established by USAID, provides nationally representative data essential for health and population research in developing countries. With over 30 years of data collection across more than 90 countries, DHS covers topics like fertility, maternal and child health, HIV/AIDS, and more. The data are collected at various spatial resolutions, typically linked to administrative units like regions or districts, with GPS coordinates recorded at the cluster level—representing small communities or neighborhoods—and displaced up to 2 kilometers in urban areas and 5 kilometers in rural areas for privacy. This level of spatial resolution allows for granular geographic analysis, which is particularly valuable when combined with mobile phone data. By integrating DHS data with mobile phone app data, researchers can gain deeper insights into population movement, health service access, and disease spread, thus enhancing the ability to track health trends, evaluate programs, and inform policy\footnote{\url{https://dhsprogram.com/Data/}}.

\textbf{Climate data } There are various sources of data for precipitation and other weather related indicators across the world as well as air pollution and other indicators that are provided through satellite imagery and modelling and assimilation projects that rely on raw satellite data combined with other data sources. Some resources include NASA's Global Precipitation Measurement\footnote{\url{https://gpm.nasa.gov/data}}, NASA's Modern-Era Retrospective analysis for Research and Applications, Version 2 (MERRA-2)\footnote{\url{https://gmao.gsfc.nasa.gov/reanalysis/MERRA-2/}}, and the European Union's Copernicus Air Monitoring System (CAMS)\footnote{\url{https://atmosphere.copernicus.eu/}}.

\textbf{World Bank Microdata Repository } In addition to making various data available on indicators like GDP, population and other country statistics\footnote{\url{https://data.worldbank.org/}}, the World Bank also has a Microdata Catalog where it can be possible to find many different microdatasets for each country. Many of these are available openly or through a data request\footnote{\url{https://microdata.worldbank.org/index.php/home}}.

\section{Concluding Remarks}

The Netmob 2024 Data Challenge dataset provides researchers with the opportunity to study mobility and population patterns for four countries across the world, with the goal of increasing knowledge and research that is especially relevant for low- and middle-income countries. Different levels of spatial and temporal aggregations can be used for different applications and studies, and there can also be important questions looked at related to the data itself and methods for tackling some of the challenges that can arise when working with passively collected data. This paper helps to present the dataset and some of these challenges. It is our hope that this reference dataset will foster the production of new research methods and the reproducibility of research outcomes.

\section*{Acknowledgements}
We would like to thank Cuebiq for providing the data and secure platform that were used to produce the final aggregated datasets for the Data Challenge. In particular, we would like to thank Brennan Lake and Éadaoin Ilten for support in working with the data to produce the datasets and reviewing research proposals. We would also like to thank the Government of Spain for funding support provided to the Global Data Facility - Mobile Phone Data for Policy Program, which is helping to sponsor Netmob 2024.

\bibliography{Netmob_Challenge_Data_Paper.bib}

\section{Appendix:}

Figure \ref{fig:heatmaps}(a) presents the OD matrix by geohash 5 for 4 countries. All trips within the same spatial unit have been removed for the figure since it is dominating. Only the top 30 pairs with the most trips are included in the visualisation. 

\begin{figure}[H]
    \centering
    \begin{subfigure}[b]{0.45\textwidth}
        \centering
        \includegraphics[width=\textwidth]{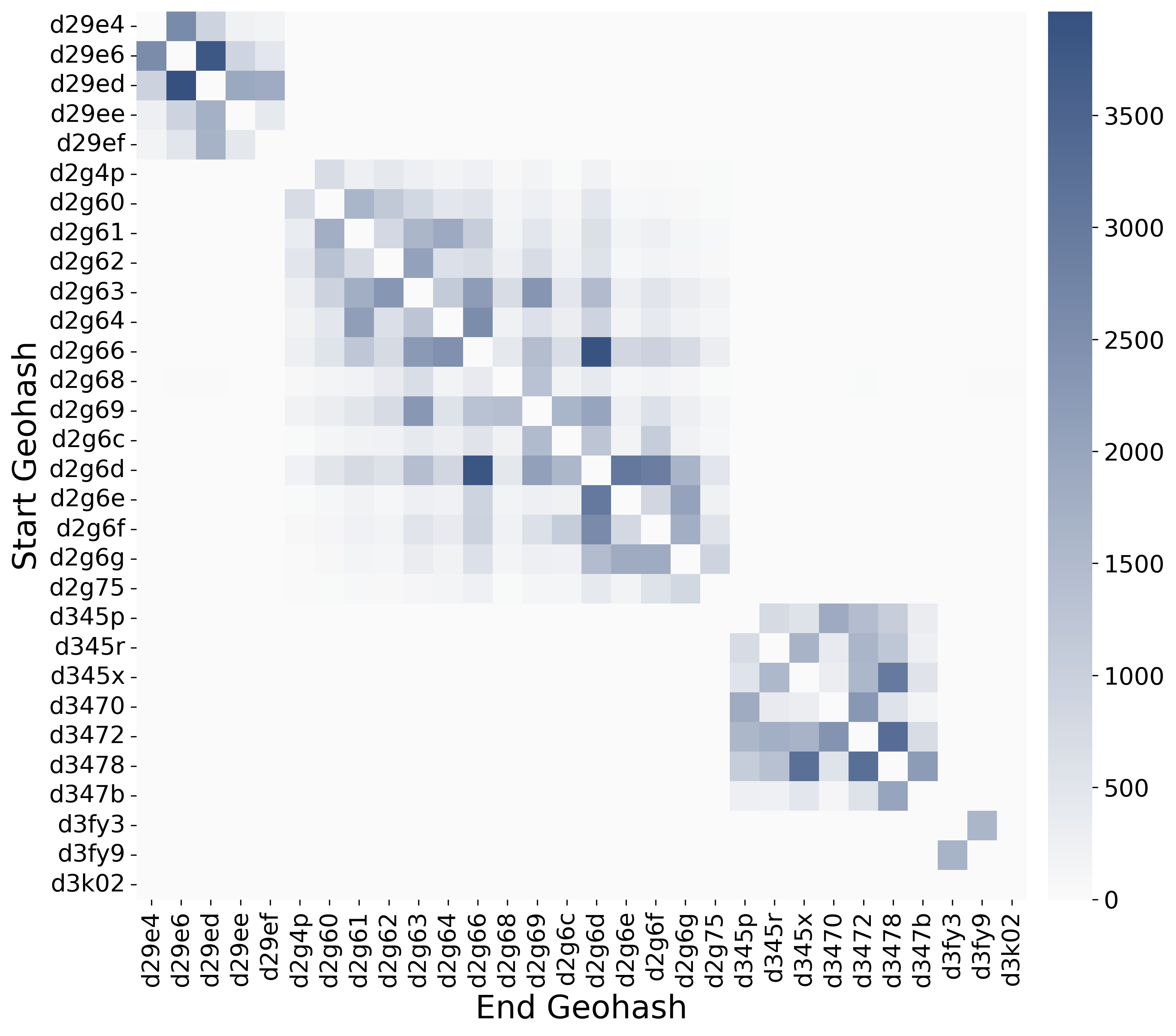} 
        \caption{\small CO}
        \label{fig:heatmap_co}
    \end{subfigure}
    \begin{subfigure}[b]{0.45\textwidth}
        \centering
        \includegraphics[width=\textwidth]{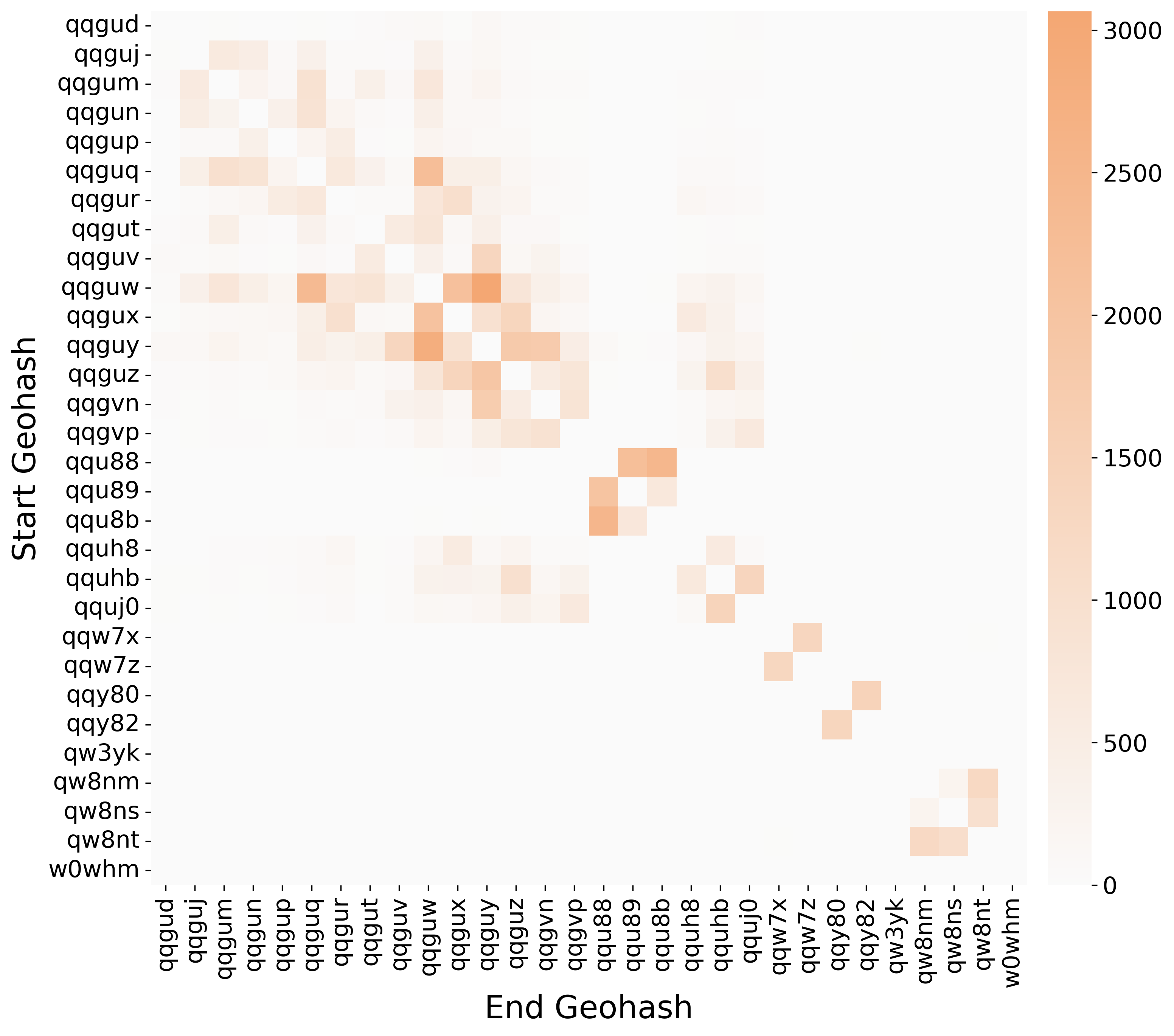} 
        \caption{\small ID}
        \label{fig:heatmap_id}
    \end{subfigure}
    \begin{subfigure}[b]{0.45\textwidth}
        \centering
        \includegraphics[width=\textwidth]{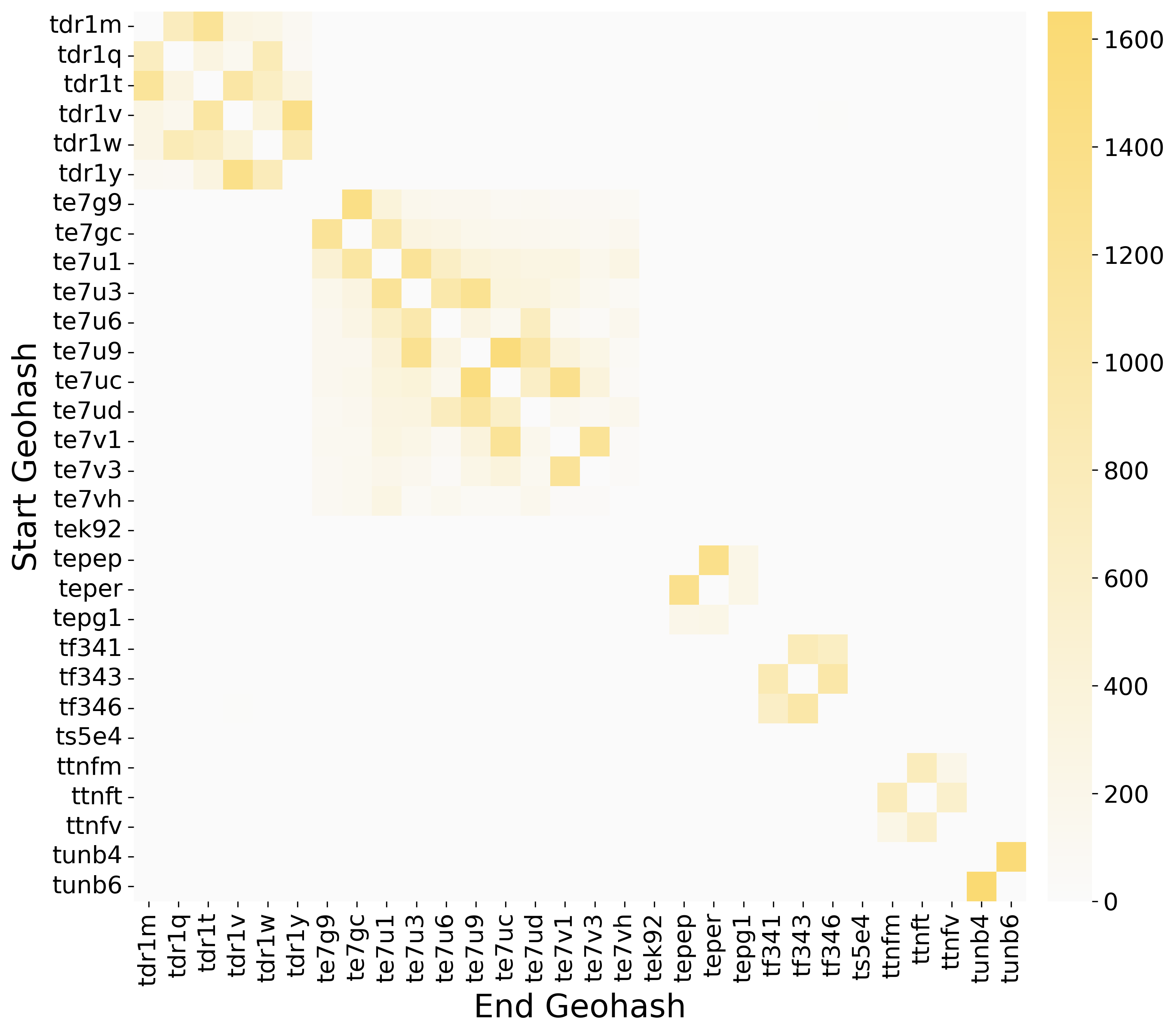}
        \caption{\small IN}
        \label{fig:heatmap_in}
    \end{subfigure}
    \begin{subfigure}[b]{0.45\textwidth}
        \centering
        \includegraphics[width=\textwidth]{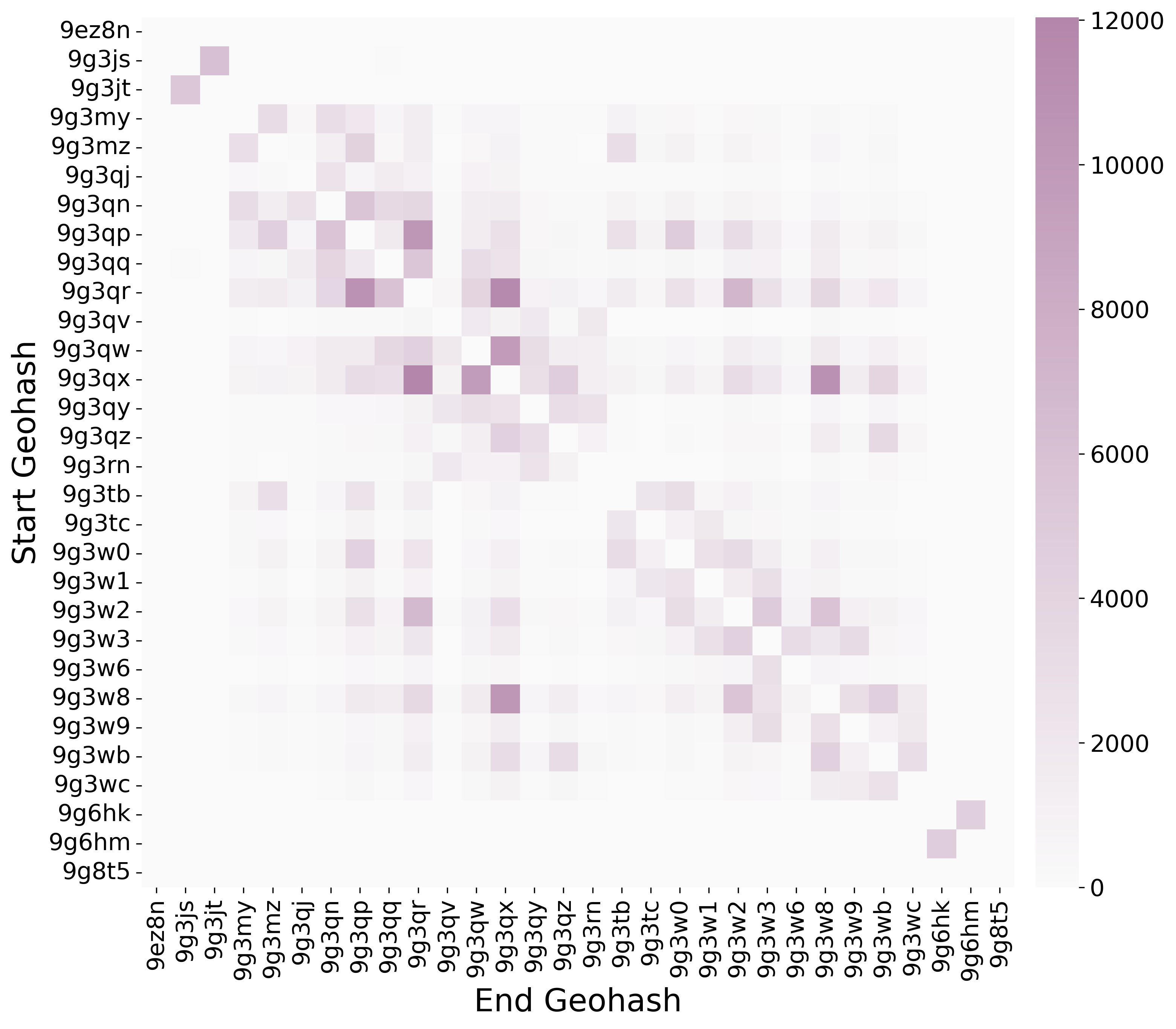} 
        \caption{\small MX}
        \label{fig:heatmap_mx}
    \end{subfigure}
    \caption{Top 30 OD Heat Map Dec 2019}
    \label{fig:heatmaps}
\end{figure}

\begin{figure}[H]
	\centering
	\includegraphics[width=1\textwidth]{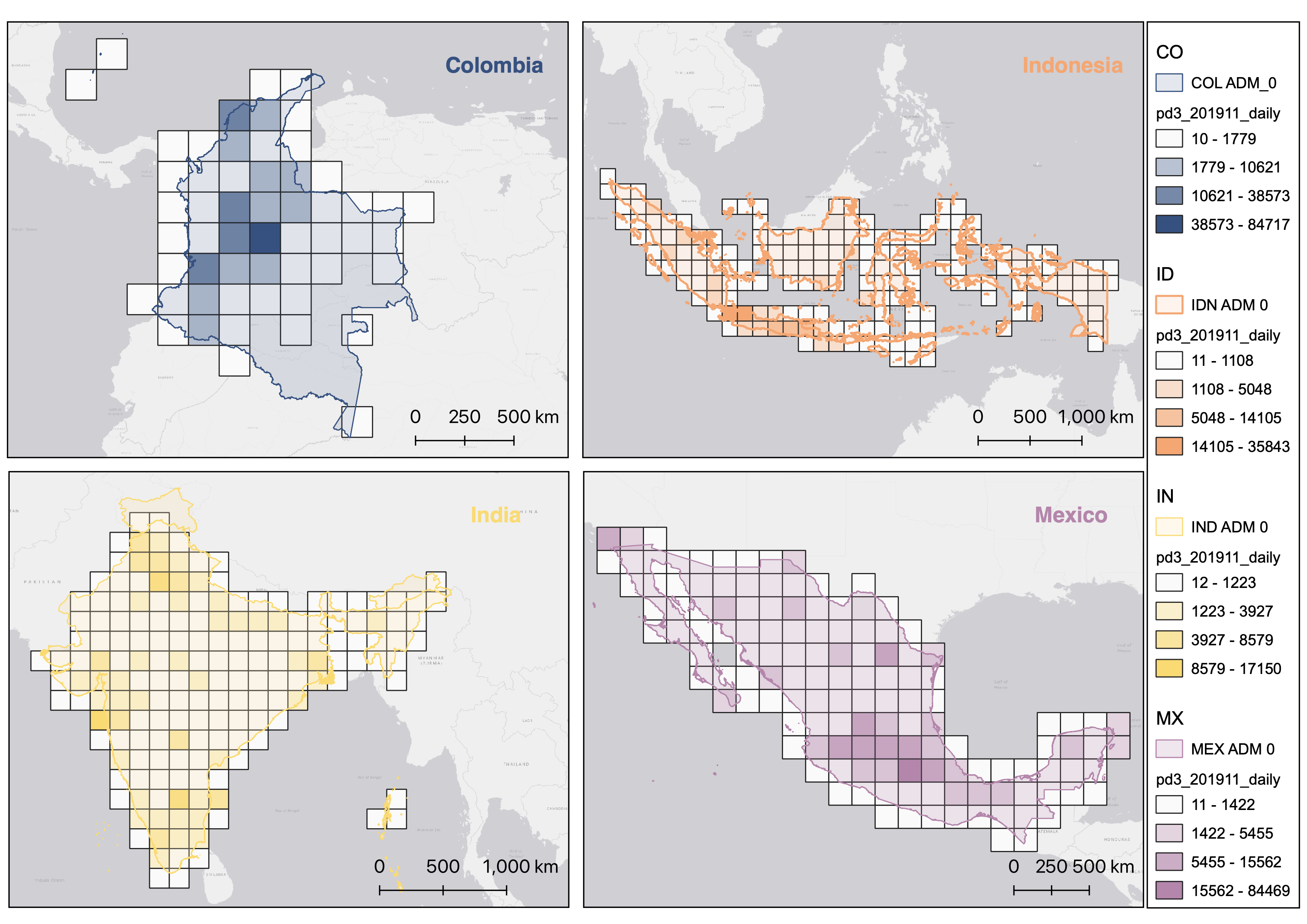}
	\caption{Unique User from Population Density Map in GH3, Average across days in Nov 2019}
	\label{fig:map}
\end{figure}

Figure \ref{fig:map} illustrates the distribution of unique users from the PD dataset at GH3 level for the four countries in the Data Challenge. The highest concentrations are observed in the capital areas of each country, with Mumbai in India also displaying a notably high population count. 

\end{document}